\documentclass{article}
\usepackage{spconf,amsmath,graphicx,hyperref}
\usepackage{amssymb}
\usepackage{booktabs}
\usepackage{graphicx}
\usepackage{subcaption}
\usepackage{float}
\usepackage{multirow}

\title{STACodec: Semantic Token Assignment for Balancing \\ Acoustic Fidelity and Semantic Information in Audio Codecs}
\name{Kaiyuan Zhang\textsuperscript{*},  Mohan Shi\textsuperscript{*}, Eray Eren, Natarajan Balaji Shankar, Zilai Wang, Abeer Alwan\thanks{\textsuperscript{*}Equal contribution.}}
\address{Dept. of Electrical and Computer Engineering, University of California, Los Angeles, USA
}

\begin{document}
\ninept
\maketitle
\begin{abstract}
Neural audio codecs are widely used for audio compression and can be integrated into token-based language models. Traditional codecs preserve acoustic details well but lack semantic information. Recent hybrid codecs attempt to incorporate semantic information through distillation, but this often degrades reconstruction performance, making it difficult to achieve both. To address this limitation, we introduce STACodec, a unified codec that integrates semantic information from self-supervised learning (SSL) models into the first layer of residual vector quantization (RVQ-1) via semantic token assignment (STA). To further eliminate reliance on SSL-based semantic tokenizers and improve efficiency during inference, we propose a semantic pre-distillation (SPD) module, which predicts semantic tokens directly for assignment to the first RVQ layer during inference. Experimental results show that STACodec outperforms existing hybrid codecs in both audio reconstruction and downstream semantic tasks, demonstrating a better balance between acoustic fidelity and semantic capability.
\end{abstract}
%
\section{Introduction}
\label{sec:intro}

Multimodal large language models \cite{DBLP:conf/iclr/WuZCTLFZXYY0025, defossez2024moshi, peng2024survey, cui2024recent} have demonstrated a powerful capability to process and generate multimodal information, including text, images, and audio. To effectively integrate speech and audio into token-based models, audio tokenization is critical. Two types of discrete audio tokens are usually employed: acoustic tokens \cite{defossez2022high, kumar2023high, parker2024scaling} and semantic tokens \cite{DBLP:conf/interspeech/ChangYFM023, DBLP:conf/icassp/ChangYCJLMSST0F24, DBLP:conf/interspeech/ShiMIS024}. Acoustic tokens, obtained from neural audio codecs, are designed for high reconstruction quality but exhibit limited semantic awareness, making them less suitable for language modeling and semantic-related tasks \cite{mousavi2024dasb, mousavi2025discrete}. Conversely, semantic tokens are typically obtained by clustering pretrained self-supervised learning (SSL) models \cite{hsu2021hubert, chen2022wavlm} or training encoders in a supervised manner \cite{du2024cosyvoice, du2024cosyvoice2}. The main drawback is that they lack fine acoustic details, limiting reconstruction quality \cite{mousavi2025discrete}. We use the term semantic tokens here to distinguish them from low-level acoustic tokens, although it is shown that semantic tokens can also capture phonetic information \cite{choi24b_interspeech}.

Recent work has focused on hybrid acoustic–semantic audio tokenizers that leverage SSL features to enhance the semantic capacity of audio codecs. In particular, SpeechTokenizer  \cite{zhang2023speechtokenizer} proposes distillation at the output of the first residual vector quantizer layer (RVQ-1), while X-Codec \cite{ye2025codec} proposes distillation at the combined output of all RVQ layers and adds SSL representations to its encoder. Besides SSL features, other approaches like PAST \cite{hartuv25_interspeech} apply extra tasks such as automatic speech recognition (ASR) and phoneme recognition (PR) to directly supervise the output of RVQ-1, so that semantic and phonetic information are built into quantization.
Although these methods achieve a hybrid tokenizer, their use of direct loss functions can degrade audio reconstruction quality. Because the training targets, whether SSL features or ASR/phoneme labels, do not correlate directly with fine acoustic details, they can drive the codebook embeddings away from accurate acoustic representations toward semantic or phonetic targets, producing lower-quality audio reconstruction compared with codecs trained only for reconstruction.
Concurrently, HASRD \cite{hussein25_interspeech} tries to disentangle semantic features from SSLs in the first codebook, avoiding the problems caused by extra supervision loss. However, it still suffers from a feature space mismatch between semantic and acoustic representations in different RVQ layers, leading to a trade-off between reconstruction quality and downstream ASR performance.

To address the limitations of previous approaches, we propose a unified codec model, STACodec, which integrates semantic information into audio codecs via semantic token assignment (STA). Unlike HASRD, which disentangles semantic features into the first codebook and causes reconstruction mismatches, our approach introduces semantic tokens (e.g., K-means tokens from SSLs) into RVQ-1 via token assignment. This design ensures accurate semantic token alignment while keeping the codebook embedding space flexible for preserving acoustic information. To eliminate reliance on SSL-based semantic tokenizers and improve efficiency during infereence, we further introduce a Semantic Pre-Distillation (SPD) module that predicts semantic tokens for assignment to RVQ-1. SPD is applied before the RVQ layers, alleviating the direct negative impact on the input to the acoustic decoder seen in previous methods. In addition, we adopt a random masking strategy during SPD to mitigate overfitting and improve distillation quality.

Experimental results show that STACodec outperforms other hybrid codecs in both audio reconstruction and representative downstream semantic-related tasks such as automatic speech recognition (ASR) and intent classification (IC), demonstrating a better balance between acoustic fidelity and semantic capability.\footnote{Code is available at \href{https://github.com/epcm/STACodec}{https://github.com/epcm/STACodec}.}

\section{Method}
\label{sec:method}

\begin{figure*}[!t]
    \centering
    \includegraphics[width=0.86\textwidth]{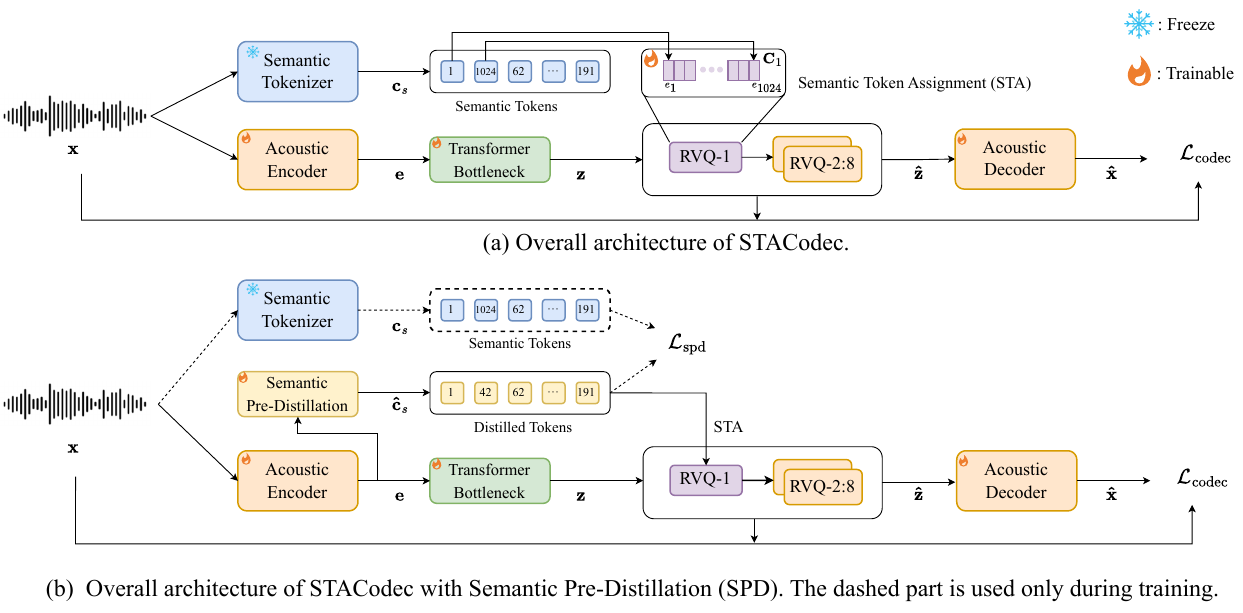}
    \caption{Overall architecture of STACodec with and without Semantic Pre-Distillation (SPD). }
    \label{fig:pipeline}
\end{figure*}\vspace{-6pt}

\subsection{STACodec with Semantic Token Assignment}
In this section, we present the detailed methodology of our proposed STACodec, as shown in Figure~\ref{fig:pipeline}(a). STACodec directly integrates semantic tokens into the first layer of the residual vector quantizer (RVQ-1) through semantic token assignment (STA), with the aim of ensuring accurate semantic token alignment while keeping the codebook embedding space flexible for preserving acoustic information. We describe the overall pipeline in Section~\ref{sec:pipeline} and the RVQ with STA in Section~\ref{sec:quant}.

\subsubsection{Overall pipeline}
\label{sec:pipeline}
We first encode the raw audio $\mathbf{x}$ into semantic tokens $\mathbf{c}_s$ using a Semantic Tokenizer (e.g., K-means on SSL features) and into latent acoustic features $\mathbf{z}$ using an Acoustic Encoder followed by a Transformer Bottleneck, as follows:
\begin{align}
&\mathbf{c}_s = \text{SemanticTokenizer}(\mathbf{x}), \\
&\mathbf{e}   = \text{AcousticEncoder}(\mathbf{x}), \\
&\mathbf{z}   = \text{TransformerBottleneck}(\mathbf{e}), \label{eq:latent}
\end{align}
where $\mathbf{c}_s \in [V]^T$ is the sequence of semantic tokens with vocabulary size $V$ and length $T$, and $\mathbf{z}$ represents the latent acoustic features.

We then quantize $\mathbf{z}$ using RVQ with semantic token assignment (STA):  
\begin{equation}
\hat{\mathbf{z}}= \text{RVQ-STA}(\mathbf{z}, \mathbf{c}_s), \label{eq:quantizer}    
\end{equation}
Details of RVQ-STA are provided in Section~\ref{sec:quant}.  
The final reconstructed audio $\hat{\mathbf{x}}$ is obtained by decoding the quantized output $\hat{\mathbf{z}}$:  
\begin{equation}
    \hat{\mathbf{x}}= \text{AcousticDecoder}(\hat{\mathbf{z}})
\end{equation}

\subsubsection{RVQ with Semantic Token Assignment}
\label{sec:quant}

During quantization, we perform RVQ with semantic token assignment (RVQ-STA). 
In RVQ-STA, the first-layer code index at time step $t$ is assigned to be the semantic token $c_{s,t}\in[V]$:
\begin{equation}
    c_{1, t} = c_{s, t}
\end{equation}

The first-layer quantized output $\mathbf{\hat{z}}_{1, t}$ is then obtained via a lookup of code index $c_{1, t}$ in the codebook $\mathbf{C}_1$, and the corresponding residual is computed:
\begin{align}
    \mathbf{\hat{z}}_{1, t} &= \mathbf{C}_1[c_{1, t}] \\
    \mathbf{r}_{1, t} &= \mathbf{z}_t - \mathbf{\hat{z}}_{1, t}
\end{align}

For the remaining layers ($i=2,\dots,N_q$), we perform standard residual vector quantization:
\begin{align}
    \mathbf{\hat{z}}_{i,t}, c_{i, t} &= \text{VQ}(\mathbf{r}_{i-1, t}; \mathbf{C}_i) \\
    \mathbf{r}_{i, t} &= \mathbf{r}_{i-1, t} - \mathbf{\hat{z}}_{i, t}
\end{align}

The final quantized vector is obtained by summing the outputs of all $N_q$ layers:
\begin{equation}
    \mathbf{\hat{z}}_{t} = \sum_{i=1}^{N_q} \mathbf{\hat{z}}_{i, t}
\end{equation}

\subsection{Semantic Pre-Distillation}
\label{sec:spd}
To eliminate the need for SSL-based semantic tokenizers and improve efficiency during inference, we propose a transformer-based semantic pre-distillation (SPD) module in addition to STACodec, which predicts semantic tokens for assignment to the first RVQ layer. The overall process is illustrated in Figure~\ref{fig:pipeline} (b). Unlike previous hybrid tokenizers such as SpeechTokenizer and X-Codec, which perform semantic distillation during or after quantization, our approach introduces distillation prior to quantization, thereby alleviating the direct negative impact on the input of the acoustic decoder.

To mitigate overfitting, we apply masking to the input of the SPD module along both the temporal and feature dimensions. Specifically, span masking is performed in each dimension with a specified probability, where contiguous segments of frames or feature channels are randomly masked. The distilled semantic tokens $\hat{\mathbf{c}}_s$ are then obtained as follows:
\begin{equation}
\hat{\mathbf{c}}_s = \text{SPD}(\text{Mask}(\mathbf{e}))
\end{equation}
During quantization (Eq.~\ref{eq:quantizer}), the original semantic tokens $\mathbf{c}_s$ are replaced with the distilled tokens $\hat{\mathbf{c}}_s$.

\subsection{Training Objective}
\label{ssec:train_obj}
For the STACodec without semantic pre-distillation (SPD), we adopt the same training objective as EnCodec \cite{defossez2022high}, which combines a reconstruction loss, a perceptual loss (via discriminators), and an RVQ commitment loss. We denote the overall objective as $\mathcal{L}_{\text{codec}}$.

For the STACodec with SPD, we introduce an additional cross-entropy loss to guide the prediction of semantic tokens from the SPD module:
\begin{equation}
\mathcal{L}_{\text{spd}} = \text{CrossEntropy}(\hat{\mathbf{c}}_{s}, \mathbf{c}_{s}),
\end{equation}
where $\hat{\mathbf{c}}_{s}$ are the semantic tokens predicted by SPD and $\mathbf{c}_{s}$ are the ground-truth tokens. The overall training objective becomes
\begin{equation}
\mathcal{L} = \mathcal{L}_{\text{codec}} + \lambda \mathcal{L}_{\text{spd}},
\end{equation}
with $\lambda$ controlling the relative weight of SPD. To ensure stable optimization, STACodec-SPD is trained in two stages: first with $\mathcal{L}_{\text{codec}}$ alone to establish reconstruction ability, and then with both $\mathcal{L}_{\text{codec}}$ and $\mathcal{L}_{\text{spd}}$, where quantization uses the distilled tokens, enabling joint optimization of semantic distillation and audio reconstruction.

\section{Experimental Setup}
\label{sec:pagestyle}

\subsection{Training Dataset}
\label{ssec:dataset}
We use LibriSpeech \cite{panayotov2015librispeech} for codec training, yielding about 960 hours of audio. To form batches during training, we randomly extract a 3-second segment for each utterance, following \cite{zhang2023speechtokenizer}.

\subsection{Model Configuration}
\label{ssec:model_config}

\textbf{Audio Codec}. The audio encoder and decoder architectures follow the configuration of EnCodec \cite{defossez2022high}, employing temporal downscaling factors of $[8, 5, 4, 2]$, which yield a frame rate of 50 Hz. The latent dimensionality is set to $D = 128$. The Transformer Bottleneck consists of 8 layers with a hidden size of 768, 16 attention heads, and a feed-forward dimension of 2048, following \cite{hartuv25_interspeech}. The RVQ component employs 8 codebooks, each containing 1024 entries. \\
\textbf{Semantic Tokenizer}. We employ K-means clustering on representations from WavLM-large (layer 23) \cite{chen2022wavlm} and HuBERT-base (layer 9) \cite{hsu2021hubert} for tokenization, as these layers have been shown to capture rich semantic information in previous works \cite{chen2022wavlm, mousavi24_interspeech}. For discrete WavLM-large, we adopt the model\footnote{\url{https://huggingface.co/speechbrain/hifigan-wavlm-k1000-LibriTTS}} from \cite{mousavi24_interspeech}, where the K-means model was trained on LibriSpeech \cite{panayotov2015librispeech} with 1000 clusters. For discrete HuBERT-base, we train a K-means model on LibriSpeech with 1024 clusters. If the tokenizer vocabulary is smaller than the codebook, only the initial codebook entries are used.\\
\textbf{Semantic Pre-Distillation}. The token distillation module adopts the same architectural configuration as the bottleneck transformer. The loss weight for $\mathcal{L}_{\text{spd}}$ is set to $\lambda = 5$, while the reconstruction loss weights follow the configuration specified in \cite{defossez2022high}. For input feature masking, we apply masks along both the temporal dimension and the feature dimension, each with a probability of 0.5. Along the temporal dimension, we mask two spans of 10 frames each out of a total of 150 frames (3 seconds). Along the feature dimension, we mask two spans of length 8 each out of $D=128$ feature dimensions.

\subsection{Training Configuration}
\label{ssec:train_config}

Training is performed on a single NVIDIA A6000 GPU with a batch size of 32. For the standard STACodec, each model is trained for approximately 280,000 steps. For STACodec with SPD, we employ a two-stage training strategy as mentioned in Section ~\ref{ssec:train_obj}, with a total of 250,000 steps: 90,000 in the first stage and 160,000 in the second. We use the Adam optimizer with betas $(0.5, 0.9)$ and no weight decay. The learning rate follows a cosine decay schedule, starting from $3 \times 10^{-4}$ with 4000 warm-up steps and gradually annealed to zero.

\subsection{Evaluation}
\label{ssec:subhead}
\subsubsection{Audio Reconstruction Evaluation}
We evaluate audio reconstruction fidelity on the test-clean subset of LibriSpeech using the VERSA toolkit \cite{shi2025versa}. The following objective metrics are reported: \textbf{PESQ} (Perceptual Evaluation of Speech Quality) \cite{rix2001perceptual}, \textbf{STOI} (Short-Time Objective Intelligibility) \cite{taal2011algorithm}, \textbf{ViSQOL} (Virtual Speech Quality Objective Listener)\cite{hines2015visqol} and \textbf{Speaker Similarity} using RawNet3 \cite{jung2022pushing} embeddings.

\subsubsection{Downstream Semantic-related tasks Evaluation}
We evaluate the semantic information preserved in tokens through two downstream tasks: Automatic Speech Recognition (ASR) and Intent Classification (IC), using the DASB benchmark \cite{mousavi2024dasb}. For ASR, we adopt the DASB setup with a two-layer BiLSTM followed by a linear projection to character sequences, trained on LibriSpeech train-clean-100 and evaluated on test-clean and test-other. This matches the model and dataset configuration used for ASR evaluation in HASRD \cite{hussein25_interspeech}. For IC, the model is a two-layer BiLSTM followed by a linear classifier, with accuracy as the metric, trained and evaluated on the SLURP dataset \cite{bastianelli2020slurp}, which contains about 72k single-turn user–assistant interactions.

\section{Experimental Results}
\label{sec:result}

\begin{table*}[ht]
\centering
\vspace{-6pt}
\caption{Evaluation of audio reconstruction and downstream semantic-related tasks. Semantic Info. denotes the source of semantic information for hybrid tokenizers. Clean and Other denote downstream ASR results trained on LibriSpeech train-clean-100 and evaluated on test-clean and test-other subsets. * indicates statistical significance ($p<0.05$) relative to the best results of open-source hybrid tokenizers.}
\vspace{-0.2cm}
\label{tab:all_eval}
\resizebox{0.85\textwidth}{!}{
\begin{tabular}{l c l c c c c c c c}
\toprule
\multirow{4}{*}{Method} & \multirow{5}{*}{\shortstack{Bitrate\\(kbps)}} & \multirow{4}{*}{Semantic Info.}
& \multicolumn{4}{c}{Audio Reconstruction} 
& \multicolumn{3}{c}{Downstream Semantic-related Tasks} \\
\cmidrule(lr){4-7} \cmidrule(lr){8-10}
& & & \multirow{2}{*}{PESQ $\uparrow$} & \multirow{2}{*}{STOI $\uparrow$} & \multirow{2}{*}{ViSQOL $\uparrow$} & \multirow{2}{*}{Spk. Sim. $\uparrow$} 
  & \multicolumn{2}{c}{ASR-WER (\%)} $\downarrow$ & \multirow{2}{*}{IC-Acc. (\%) $\uparrow$} \\
\cmidrule(lr){8-9}
& & & & & & & Clean  & Other & \\
\midrule
Ground Truth          & - & \multicolumn{1}{c}{-}          & 4.64 & 1.00 & 5.00 & 1.00 & - & - & - \\
\midrule
SpeechTokenizer\cite{zhang2023speechtokenizer}      & 4.0 & HuBERT-base & 2.60 & 0.92 & 4.26 & 0.85 & 18.63 & 41.87 & 56.61 \\
X-Codec\cite{ye2025codec}               & 4.0 & HuBERT-base & 2.79 & 0.88 & 4.27 & 0.88 & 16.48 & 37.22 & 66.49 \\
PAST\cite{hartuv25_interspeech}                 & 4.0 & ASR and PR  & 3.16 & 0.96 & 4.32 & 0.86 & 15.83 & 37.02 & 59.50 \\
\multirow{2}{*}{HASRD\cite{hussein25_interspeech}}                
& 4.5 & HuBERT-base     & -    & -    & 4.30 & -    & 11.30 & -     & - \\
& 3.1 & BestRQ+     & -    & -    & 4.50 & -    & 21.00 & -     & - \\
\midrule
\multirow{2}{*}{STACodec}               & 4.0 & HuBERT-base & 3.61 & 0.97 & 4.50 & 0.93 & 10.94 & 23.54 & 70.81 \\
                      & 4.0 & WavLM-large & \textbf{3.62*} & \textbf{0.97*} & \textbf{4.51*} & \textbf{0.93*} 
                                     & \textbf{9.35*} & \textbf{21.89*} & \textbf{74.21*} \\
\midrule
STACodec-SPD  & 4.0 & WavLM-large & 3.51 & 0.96 & 4.43 & 0.92 & 15.39 & 35.05 & 64.31 \\
\bottomrule
\end{tabular} 
}
\vspace{-8pt}
\end{table*}

\subsection{Comparison with Baselines}
We compare STACodec with four recent hybrid tokenizers: SpeechTokenizer \cite{zhang2023speechtokenizer}, X-Codec \cite{ye2025codec}, PAST \cite{hartuv25_interspeech}, and HASRD \cite{hussein25_interspeech}, which include SOTA hybrid tokenizers. For HASRD, we use the results reported in the paper because it is not open-sourced. For other methods we use the official checkpoints. The overall comparison is summarized in Table~\ref{tab:all_eval}. 

Compared with baselines that use HuBERT-base for semantic information, STACodec achieves significantly better audio reconstruction quality as well as better performance on downstream semantic-related tasks. For reconstruction, PESQ improves from 2.79 (X-Codec) to 3.61, and ViSQOL from 4.30 (HASRD) to 4.50, showing that the proposed semantic token assignment (STA) causes less degradation in preserving acoustic details than other semantic integration methods. For downstream tasks, STACodec attains a lower ASR WER than HASRD (10.94\% vs. 11.30\%) and improves IC accuracy from 66.49\% (X-Codec) to 70.81\%, demonstrating that STA provides a more effective way to preserve rich semantic information in tokens.

Replacing the HuBERT-base K-means semantic tokenizer with WavLM-large K-means yields better results in both audio reconstruction and downstream semantic-related tasks, outperforming the baselines on all metrics. Interestingly, HASRD exhibits a clear trade-off between reconstruction quality and semantic performance: switching from HuBERT-base to BestRQ+ \cite{hussein25_interspeech} improves ViSQOL from 4.30 to 4.50 but leads to a substantial drop in ASR performance, with WER increasing from 11.30\% to 21.00\%. This suggests that the mismatch between semantic and acoustic features underlies this trade-off. When the semantic features disentangled in the first VQ layer are over-optimized to resemble fine acoustic features, reconstruction quality improves but semantic information is lost, and the reverse holds true as well.

STACodec with semantic pre-distillation (SPD) eliminates the need for an SSL-based semantic tokenizer during inference, reducing the inference parameter count by 250M and 30 GFLOPs per second of audio when compared with STACodec that uses WavLM-large K-means as the semantic tokenizer, while maintaining strong performance on both reconstruction and downstream semantic tasks. Compared with SpeechTokenizer and X-Codec, which apply distillation either during or after quantization, STACodec-SPD delivers substantially better reconstruction quality, showing that the proposed pre-distillation strategy alleviates the degradation of audio reconstruction typically caused by distillation.  Compared with PAST, which directly supervises specific downstream tasks such as ASR and PR, STACodec-SPD achieves slightly better ASR and considerably higher IC accuracy. These results demonstrate the effectiveness of the proposed SPD.

\subsection{Codebook Utilization}
\begin{figure}[t]
    \centering
    \includegraphics[width=\linewidth]{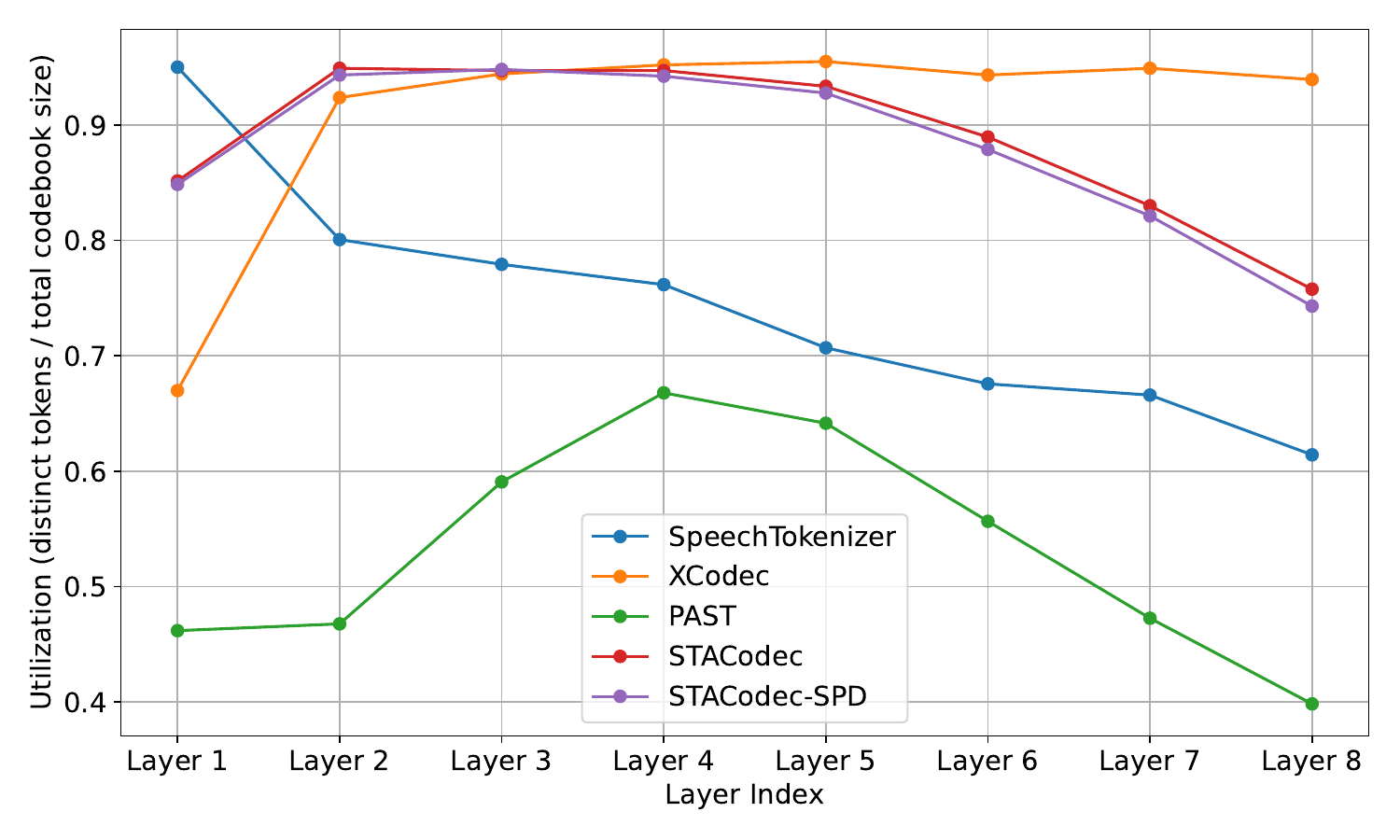} 
    \vspace{-0.5cm}
    \caption{Comparison of codebook utilization across RVQ layers for different hybrid tokenizers, calculated on 1,000 utterances from the LibriSpeech test-clean subset.}
    \label{fig:codebook}
    \vspace{-0.4cm}
\end{figure}
To investigate why STACodec achieves an effective balance between acoustic and semantic information preservation, Figure~\ref{fig:codebook} presents the codebook utilization for various hybrid tokenizers, computed from 1,000 samples of the LibriSpeech test-clean set. Hybrid tokenizers that impose direct supervision on the output of the first VQ layer (e.g., SpeechTokenizer and PAST) exhibit low codebook utilization in layers 2–8. In particular, PAST, which directly supervises ASR and PR, shows especially low utilization across all RVQ layers. X-Codec, which reconstructs semantic features after all RVQ layers, continues to show low utilization in the first layer. In contrast, STACodec and its SPD variant maintain balanced codebook utilization across layers, indicating more efficient use of the feature space.

\subsection{Ablation Experiments}
\begin{table}[t]
\centering
\caption{Ablation experiments of STACodec components. STA denotes semantic token assignment, TC denotes trainable codebook for RVQ-1, BT denotes the transformer bottleneck, and Mask denotes input masking used for training the SPD module. PESQ evaluates audio reconstruction quality, while ASR-WER (\%) measures downstream ASR performance, trained on LibriSpeech train-clean-100 and evaluated on test-clean.}
\vspace{-0.1cm}
\label{tab:ablation}
\resizebox{0.85\columnwidth}{!}{
\begin{tabular}{c c c c c c c}
\toprule
 & STA & BT & TC & Mask & PESQ $\uparrow$ & ASR-WER (\%) $\downarrow$ \\
\midrule
\multirow{3}{*}{w/o SPD} 
  & $\times$ & $\times$ & \checkmark & - & 3.88 & 40.62 \\
  & \checkmark & $\times$ & \checkmark & - & 3.58 & 9.27 \\
  & \checkmark & \checkmark & $\times$ & - & 3.46 & 9.33 \\
  & \checkmark & \checkmark & \checkmark & - & 3.62 & 9.35 \\

\midrule
\multirow{2}{*}{w/ SPD} 
  & \checkmark & \checkmark & \checkmark & $\times$ & 3.54 & 15.70 \\
  & \checkmark & \checkmark & \checkmark & \checkmark & 3.51 & 15.39 \\
\bottomrule
\end{tabular}
}
\vspace{-0.4cm}
\end{table}

Table~\ref{tab:ablation} presents the ablation experiments on the individual components of STACodec and SPD. PESQ and ASR-WER are used as representative metrics for reconstruction quality and downstream performance, respectively. For STACodec without SPD, the proposed STA method significantly reduces ASR-WER, indicating its effectiveness in preserving semantic information. The design of trainable codebook in RVQ-1 can significantly improve reconstruction quality. The inclusion of the transformer bottleneck improves reconstruction quality, with a slight degradation in ASR. For STACodec with SPD, introducing masking enhances the effectiveness of tokenizer distillation and yields better downstream ASR performance. However, this comes at the cost of reconstruction fidelity, highlighting the inherent trade-off between preserving semantic and acoustic information.


\vspace{-0.3cm}
\section{Conclusion}
\label{sec:conclusion}
\vspace{-0.1cm}
In this work, we propose a novel unified codec model, STACodec, which seamlessly integrates semantic information from self-supervised learning (SSL) models into audio codecs via semantic token assignment (STA), while keeping the codebook embedding space flexible for preserving acoustic information. In addition, we introduce semantic pre-distillation (SPD), which directly predicts semantic tokens for assignment to the first RVQ layer during inference, thereby eliminating the need for additional SSL inference and clustering. Experiments show that STACodec outperforms existing hybrid tokenizers in both audio reconstruction and downstream semantic-related tasks such as ASR and IC. Moreover, STACodec with SPD reduces the inference parameter count introduced by SSL-based semantic tokenizers while maintaining reasonable reconstruction and downstream performance. 

\section{Acknowledgments}The research is supported in part by the IES, U.S. Department of Education (DoE), through Grant R305C240046 to the U. of Buffalo. The opinions expressed are those of the authors and do not represent views of the IES and DoE.

\bibliographystyle{IEEEbib}
\bibliography{strings,refs}

\end{document}